
\magnification=\magstep1
\hsize=128mm
\baselineskip=18pt
\vglue 1in

\centerline{\bf UNIVERSAL PARAMETRIC CORRELATIONS AT THE SOFT EDGE}
\centerline{\bf OF THE SPECTRUM OF RANDOM MATRIX ENSEMBLES}
\vskip 10mm
\centerline{A. M. S. Mac\^edo }
\centerline{\it Theoretical Physics, University of Oxford,}
\centerline{\it 1 Keble Road, Oxford OX1 3NP, U.K.}
\vskip 10mm
\centerline{\bf Abstract}
\medskip
We extend a recent theory of parametric correlations in the spectrum of random
matrices to study the response to an external perturbation of eigenvalues near
the soft edge of the support. We demonstrate by explicit non-perturbative
calculation that, for systems with unitary symmetry, the two-point function for
level density fluctuations becomes, after appropriate rescaling, a universal
expression.
\medskip



\vfill\eject
Random matrix theory (RMT) [1] has recently been the focus of much attention
because of its wide range of applications in many branches of physics, such as
complex nuclei [2], disordered metallic grains [3], quantum chaotic systems
[4], random surfaces [5], disordered conductors [6], QCD [7] and more recently
strongly correlated many-particle systems [8]. One of the most striking and
robust predictions of RMT is the celebrated Wigner-Dyson statistics [9], which
states that the probability distribution of level spacings is a universal
function if the eigenvalues are measured in units of the mean level spacing.
Another important feature of RMT is that all physical systems are classified
according to its symmetries in just three universality classes, namely:
unitary, orthogonal and symplectic.

A significant extension of RMT has recently been put forward by Simons,
Altshuler, Lee and Szafer [10-11]. In particular, they have shown that in order
to describe the statistical properties of the dispersion of the energy levels
of a particle diffusing through a disordered metallic grain in response to an
external adiabatic perturbation, one has to introduce a new parameter in the
theory in addition to the mean level spacing. This new parameter has been given
the meaning of a generalized conductance because of its relation to the rate of
energy dissipation of the system due to the presence of the external
perturbation.
 The most remarkable result of their work, however, is that the $n$-point
correlation function for density of states fluctuations becomes a system
independent {\it universal} expression if the energy levels are measured in
units of the mean level spacing and if the perturbation parameter is
renormalized by the mean square gradient of the levels. This striking
universality has been analytically derived for disordered metallic systems with
orthogonal, unitary and symplectic symmetries and has been verified numerically
in a number of chaotic [10,12] and also interacting systems [13]. More
recently, Simons, Lee and Altshuler [14] have demonstrated that the non-linear
$\sigma$-model provides a common mathematical structure in which a number of
problems in different branches of theoretical physics can be interrelated.

The universality obtained by Simons {\it et al.}, and similarly that of
Wigner-Dyson statistics, applies only for levels away from the edge of the
support of the spectrum. This statement can, in principle, be formally
justified by means of renormalization group arguments. A simple renormalization
group procedure has recently been devised by Br\'ezin and Zinn-Justin [15].
These authors have shown that there are two kinds of fixed points in the
renormalization group equations: a stable gaussian one governing the behavior
of the system at the bulk of the spectrum and an unstable one governing a small
crossover region around the endpoint of the support where the average density
goes to zero as a power law.
The attractive gaussian fixed point supports the general validity of the
Wigner-Dyson statistics everywhere at the bulk of the spectrum. It is therefore
natural to expect that the results of refs. [10-14] are of similar general
validity, although an explicit proof is not yet available.

At the edge of the spectrum, however, Wigner-Dyson statistics breaks down and
an entirely new regime of parametric correlations emerges. It is well known
that the average eigenvalue densities of random matrix ensembles exhibit two
kinds of edges: a hard edge and a soft edge, where the eigenvalue integration
range is bounded and unbounded respectively.
The hard edge of the Laguerre ensemble is important in the description of many
physical systems such as disordered metallic conductors [6] and QCD [7] and has
been the subject of a recent paper [16], in which the universality of the
parameter dependent two-point level density autocorrelator has been explicitly
demonstrated. The objective of the present work is to study the response to an
external perturbation of the eigenvalues of a random matrix near the soft edge
of the support of the spectrum.
We demonstrate explicitly that, for systems with unitary symmetry, the
two-point correlation function for level density fluctuations becomes a
universal function after rescaling the levels by an appropriate system
dependent constant and the perturbation parameter by the mean square gradient
of the levels. In order to emphasize the robustness of the final result we
consider two different kinds of unitary ensembles: the Gaussian unitary
ensemble (GUE) and the Laguerre unitary ensemble (LUE). Physically, they
describe systems with broken time-reversal symmetry.

The problem of the behavior of random matrix models near the soft edge of the
spectrum is of wide interest and has been motivated by the application of RMT
to the physics of random surfaces [5]. It has been demonstrated [17,18] that
the double scaling limit of these models, in which the coupling constant of the
theory approaches a critical value and the size of the random matrix tends to
infinity, is characterized by a universal crossover function of scaling
variables, whose definitions depend on the order of multicriticality of the
problem.
The crossover region has been shown to be around the soft edge of the support
of the spectrum and its universality is controlled by an unstable fixed point
in the renormalization group flow [15]. We remark that the two random matrix
ensembles studied in the present work have the same order of multicriticality
and therefore must exhibit the same universal behavior, as will be demonstrated
by explicit calculation.

An interesting application of the model we consider can be found in ref. [19],
in which the string theory of matter with central charge equal one coupled with
2D quantum gravity is represented as a continuous matrix model similar to the
one studied in ref. [14], where its connection to the non-linear $\sigma$-model
and the Dyson Brownian motion model [20] has been demonstrated.

In this work, however, we shall, for the sake of generality, consider as our
starting point directly the Dyson Brownian motion model rather than the
continuous matrix model of refs. [14] and [19]. The Dyson Brownian motion model
describes the evolution of the probability distribution $P^{\mu}(\{x_i\},\tau)$
of $N$ classical particles at positions $\{x_i(\tau)\}$ moving in a viscous
fluid with friction coefficient $\gamma^{-1}$ at temperature $\beta^{-1}$,
$$
\gamma{\partial P^\mu\over \partial \tau} =
\sum_{i=1}^{N} {\partial \over \partial x_i}
\left(P^\mu{\partial W^\mu\over \partial x_i}+{1\over \beta}{\partial P^\mu
\over \partial x_i}\right),
\eqno (1a)
$$
in which $\mu=\{G,L\}$ and
$$
W^G(\{x_i\})= {c } \sum_i x_i^2
-\sum_{i<j} \ln\left|x_i-x_j\right|,
\eqno (1b)
$$
for the gaussian ensemble [21,22], where $-\infty \le x_i \le \infty$ and
$$
W^L(\{x_i\})=-(\alpha+{1 \over \beta})\sum_i \ln x_i + {c \over 2} \sum_i x_i^2
-\sum_{i<j} \ln\left|x_i^2-x_j^2\right|,
\eqno (1c)
$$
for the Laguerre ensemble [16], where $0 \le x_i \le \infty$. The parameters
$c$ and $\alpha$ are system dependent constants and $\beta=1,2$ and $4$ for
systems with orthogonal, unitary and symplectic symmetries.
The fictitious time $\tau$ can have several meanings depending on the
particular physical problem. It can, for instance, be the one spatial dimension
of the target space of the continuous matrix model representing the string
theory of ref. [19] or it can be related to the perturbation parameter $\delta
u$ of refs. [10-14,21] by $\tau=\delta u^2$. The average level density
$\rho^\mu(x)\equiv \langle \sum_i \delta(x-x_i(u))\rangle_{eq}$, where $\langle
\dots \rangle_{eq}$ stands for an arbitrary average calculated with the
equilibrium distribution $P^\mu(\{x_i\},\infty)$, can be calculated explicitly
by direct integration of Eq. (1). We find
$$
\rho^G(x)=\cases{ (2c^{1/2}/ \pi)\sqrt{N-cx^2} & for $|x|\le (N/c)^{1/2}$ \cr
0 & for $|x| > (N/c)^{1/2}$ \cr}
\eqno (2a)
$$
which is the Wigner semicircle law [1] and
$$
\rho^L(x)=\cases{ (c^{1/2}/ \pi)\sqrt{4N-cx^2} & for $0\le x\le 2(N/c)^{1/2}$
\cr
0 & for $x > 2(N/c)^{1/2}$ \cr},
\eqno (2b)
$$
which is a quarter circle as was first shown by Bronk [23].
Note that $\rho^\mu(x)$ has a soft edge at $x_0^G=\sqrt{N/c}$ and
$x_0^L=2\sqrt{N/c}$. Near this edge, the average level density behaves as
$\rho^\mu(x)\sim N^k$ in a region of size $\sim N^{-k}$, in which $k$ is a
multicritical exponent [18], which can be shown to be equal $1/6$ for both GUE
and LUE. The double scaling regime of these ensembles is obtained simply by
moving the origin of the spectrum to $x_0^\mu$ and measuring the eigenvalue
spacing in units of order $N^{-1/6}$. An explicit calculation of the eigenvalue
density crossover function can be found in refs. [18] and [24].

In what follows we shall be interested in studying the two-point function for
level density fluctuations defined as
$$
S^\mu(x,y,\delta u)\equiv \Bigl\langle \sum_{ij} \delta(x-x_i(u))
\delta(y-x_j(u+\delta u)) \Bigr\rangle_{eq}-\rho^\mu(x)\rho^\mu(y)
\eqno (3)
$$
in the limit $x \to x^\mu_0$ and $y \to x^\mu_0$.

It has been shown [11,16,21,22] that Eq. (1) can be exactly mapped onto a
Schr\"odinger equation in imaginary time. The case $\beta=2$, for systems with
unitary symmetry, is particularly simple since the two-body interaction term
vanishes and we end up with a theory of free fermions. In second quantization
the resulting Hamiltonian reads
$$
{\cal H}^\mu=\sum_{p=0}^{\infty} \varepsilon_p^\mu c^{\dagger}_pc^{}_p,
\eqno (4)
$$
where $\varepsilon^G_p=(2c/\gamma)(p+1/2)$,
$\varepsilon^L_p=(2c/\gamma)(p+(\alpha+1)/2)$ and $c^{\dagger}_p$  $(c^{}_p)$
creates (annihilates) a fermion with quantum number $p$. It is convenient at
this stage to introduce the field operators $\psi^{}_\mu(x)$ and
$\psi^{\dagger}_\mu(x)$, where $\psi^{}_\mu(x)\equiv \sum_p
\phi^\mu_p(x)c_p^{}$ and $\psi^{\dagger}_\mu(x)\equiv \sum_p
\phi^\mu_p(x)c_p^{\dagger}$, in which
$$
\phi^G_p(x)=\left({2c \over \pi}\right)^{1/4}(2^p p!)^{-1/2}e^{-cx^2}
H_p(x\sqrt{2c})
\eqno (5a)
$$
and
$$
\phi^L_p(x)=\left({2c^{1+\alpha} p! \over \Gamma(p+1+\alpha)}\right)^{1/2}
x^{\alpha+1/2}e^{-c x^2/2}L^\alpha_p(c x^2).
\eqno (5b)
$$
The functions $H_p(x)$ and $L^\alpha_p(x)$ are the conventional Hermite and
associate Laguerre polynomials respectively. Since ${\cal H}^\mu$ is a free
electron Hamiltonian we can use Wick's theorem to evaluate $S^\mu(x,y,\delta
u)$. We find
$$
S^\mu(x,y,\delta u)=\sum_{p=0}^\infty
\phi^\mu_p(x)\phi^\mu_p(y)\sum_{q=0}^{N-1}
\phi^\mu_q(x)\phi^\mu_q(y)\exp(2c\delta u^2 (q-p)/\gamma).
\eqno (6)
$$
For large $N$ and $x\to x_0^\mu$ we can use the asymptotics [25]
$$
e^{-x^2/2}H_N(x) \simeq
\pi^{1/4}2^{N/2+1/4}(N!)^{1/2}N^{-1/12}Ai((x-\sqrt{2N+1})N^{1/6}2^{1/2})
\eqno (7a)
$$
and
$$
e^{-x^2/2}L_N^\alpha(x^2)\simeq
(-1)^N2^{-\alpha -1/3}N^{-1/3}Ai((x-2N^{1/2})2^{2/3}N^{1/6}),
\eqno (7b)
$$
in which $Ai(x)$ is the Airy function. Inserting (7) and (5) into (6) and
taking $N \to \infty$ and $c \to 0$, such that $2c^{1/2}N^{1/6} \to \rho_0^G$
and $2^{2/3}c^{1/2}N^{1/6} \to \rho_0^L$ we get
$$\eqalignno{
S^\mu(X,Y,\delta u)=&{\rho_0^\mu}^2\int_0^\infty dz\int_0^\infty dz'
Ai(\rho_0^\mu X -z)Ai(\rho_0^\mu Y -z)Ai(\rho_0^\mu X +z') \cr
&Ai(\rho_0^\mu Y +z')\exp(-{\delta u^2 {\rho_0^\mu}^2 \over 2 \gamma}
(z+z')), & (8a) \cr
}
$$
where $X\equiv x-x_0^\mu$ and $Y\equiv y-y_0^\mu$. The parameter $\rho_0^\mu$
is related to the average level density through $\rho^\mu(x)=\rho_0^\mu
\int_0^\infty dz {Ai}^2(\rho_0^\mu x +z)$. Note that if we make the rescalings
[10-14] $\hat X=\rho_0^\mu X$, $\hat Y=\rho_0^\mu Y$, $\delta \hat u^2=\delta
u^2 {\rho_0^\mu}^2/\gamma$ and $\hat S(\hat X,\hat Y,\delta \hat
u)=S^\mu(X,Y,\delta u)/{\rho_0^\mu}^2$ we find that
$$\eqalignno{
\hat S(\hat X,\hat Y,\delta \hat u)=&\int_0^\infty dz\int_0^\infty dz'
Ai(\hat X -z)Ai(\hat Y -z)Ai(\hat X +z') \cr
&Ai(\hat Y +z')\exp(-{\delta \hat u^2  \over 2 }
(z+z')) & (8b) \cr
}
$$
becomes a function that does not depend on any physical parameter of the system
and therefore is a universal crossover function characterizing the response to
an arbitrary external perturbation of the eigenvalues of GUE and LUE near the
soft edge of the support of the spectrum. This is the central result of the
present paper.

We can easily show that Eq. (8a) reproduces known results for $\delta u=0$ if
we use the relation [26] $\int_{-\infty}^\infty dz Ai(x+z)Ai(y+z)=\delta(x-y)$
to write $S^\mu(x,y,0)$ as
$$
S^\mu(x,y,0)=\delta(x-y)\rho^\mu(x)-(K^\mu(x,y))^2,
\eqno (9a)
$$
where
$$
K^\mu(x,y)=(Ai(\rho_0^\mu x)Ai'(\rho_0^\mu y)-Ai(\rho_0^\mu y)Ai'(\rho_0^\mu
x))/(x-y).
\eqno (9b)
$$
Equation (9b) constitutes the Airy kernel [24], which is known to characterize
the soft edge of the GUE and LUE. Equation (9a) is also well known in RMT and
is usually derived using the orthogonal polynomial technique [1].

An important application of Eq. (8a) is to calculate the density of level
gradients defined as $F(v)\equiv \langle \delta(v-dx/du) \rangle$. One can show
that [27]
$$
F(v)=\lim_{\delta u \to 0} {\delta u \over \rho^\mu(x)} S^{\mu}(x,x+v \delta
u,\delta u).
\eqno (10)
$$
Asymptotic evaluation of (8a) for $\delta u \ll 1$ yields
$$
S^\mu(x,y,\delta u) \simeq \left({\gamma \over 2 \pi \delta u^2}\right)^{1/2}
\exp\left(-{\gamma(x-y)^2 \over 2 \delta u^2}\right)K^\mu(x,y).
\eqno (11)
$$
Inserting (11) into (10) gives
$$
F(v)=\left({\gamma \over 2\pi}\right)^{1/2}\exp\left(-{\gamma v^2 \over
2}\right).
\eqno (12)
$$
This result is known to be valid at the bulk of the GUE and at the bulk and
hard edge of the LUE. The present result, valid at the soft edge of both the
GUE and LUE, leads to the striking conclusion that the density of level
gradients is gaussian everywhere in the spectrum. This allow us to introduce,
also at the edge of the spectrum, the concept of a generalized conductance
characterizing the rate of energy dissipation of the system in response to the
external perturbation. It is defined as [10]
$$
C_0\equiv {\rho_0^\mu}^2\langle v^2 \rangle={\rho_0^\mu}^2/\gamma.
\eqno (13)
$$
If we now compare (13) with the rescaling of the perturbation parameter $\delta
u$ that leaves the two-point function for level density fluctuations universal,
we can see that the rescaling can be written as $\delta \hat u=C_0^{1/2} \delta
u$, which is exactly the same as that used in refs. [10-14]. The rescalings of
$X$ and $Y$ differ from that of the bulk because there is no local translation
symmetry at the edge of the spectrum.

In conclusion, we have studied the response to an external perturbation of the
eigenvalues of GUE and LUE near the soft edge of the support of the spectrum.
The main effect of the external perturbation is to cause the levels to disperse
in disjoint manifolds, as a result of eigenvalue repulsion or non-integrability
of the underlying dynamics. We have shown that the two-point level density
autocorrelator becomes a universal function when the levels are measured in
units of an appropriate system dependent constant and the perturbation
parameter is rescaled by the mean square gradient of the levels. The resulting
universal expression is identical for GUE and LUE and describes the crossover
region where the average level density goes to zero as a power law. In
addition, we have demonstrated that the density of level gradients is gaussian
everywhere in the spectrum.

The author would like to thank J. T. Chalker and B. D. Simons for many useful
comments and suggestions. This work was partially supported by CAPES (Brazilian
Agency).

\bigskip
\noindent
{\bf References}
\medskip
{\parindent=0.5cm
\item{[1]} For a review see Mehta M. L., {\it Random Matrices}, 2nd edition
(Academic, New York, N.Y.) 1991.
\item{[2]} See e.g. Porter C. E. (Editor) {\it Statistical Theory of Spectra:
Fluctuations} (Academic, New York, N.Y.) 1965.
\item{[3]} Gorkov L. P. and Eliashberg G. M., \v Z. \. Eksp. Teor. Fiz., {\bf
48} (1965) 1407 [Sov. Phys. JETP, {\bf 21} (1965) 940 ]; Efetov K. B., Adv.
Phys., {\bf 32} (1983) 53; Altshuler B. L. and Shklovski\u \i{} B. I., \v Z. \.
Eksp. Teor. Fiz., {\bf 91} (1986) 220 [Sov. Phys. JETP, {\bf 64} (1986) 127].
\item{[4]} See e.g. Haake F., {\it Quantum Signatures of Chaos} (Springer,
Berlin) 1992.
\item{[5]} Br\'ezin E., Itzykson C., Parisi G. and Zuber J. B., Commun. Math.
Phys., {\bf 59} (1978) 35.
\item{[6]} For a review see Stone A. D., Mello P. A., Muttalib K. A., and
Pichard J.-L., in {\it Mesoscopic Phenomena in Solids}, edited by Altshuler B.
L., Lee P. A., and Webb R. A. (North-Holland, Amsterdam)  1991.
\item{[7]} Verbaarschot J. J. M. and Zahed I., Phys. Rev. Lett., {\bf 70}
(1993) 3852.
\item{[8]} Montambaux G., Poilblanc D., Bellissard J. and Sire C., Phys. Rev.
Lett., {\bf 70} (1993) 497.
\item{[9]} Wigner E. P., Ann. Math., {\bf 53} (1951) 36; Dyson F. J., J. Math.
Phys., {\bf 3} (1962) 140.
\item{[10]} Szafer  A. and Altshuler B. L., Phys. Rev. Lett., {\bf 70} (1993)
587;  Simons B. D. and Altshuler B. L., ibid., {\bf 70} (1993) 4063; Phys. Rev.
B, {\bf 48} (1993) 5422; Simons B. D., Lee P. A. and Altshuler B. L., Phys.
Rev. B, {\bf 48} (1993) 11 450.
\item{[11]} Simons B. D., Lee P. A. and Altshuler B. L., Phys. Rev. Lett., {\bf
70} (1993) 4122.
\item{[12]} Simons B. D., Hashimoto A., Courtney M., Kleppner D. and Altshuler
B. L., Phys. Rev. Lett., {\bf 71} (1993) 2899.
\item{[13]} Faas M., Simons B. D., Zotos X. and Altshuler B. L., Phys. Rev. B,
{\bf 48} (1993) 5439.
\item{[14]} Simons B. D., Lee P. A. and Altshuler B. L., Nucl. Phys., {\bf
B409} (1993) 487; Phys. Rev. Lett. (to appear).
\item{[15]} Br\'ezin E. and Zinn-Justin J., Phys. Lett. B, {\bf 288} (1992) 54.
\item{[16]} Mac\^ edo  A. M. S. (unpublished).
\item{[17]} Br\'ezin  E. and Kazakov V., Phys. Lett. B, {\bf 236} (1990) 2125;
Douglas M. R.  and Shenker S. H., Nucl. Phys. B, {\bf 335} (1990) 635; Gross D.
J. and Migdal A. A., Phys. Rev. Lett., {\bf 64} (1990) 27;
\item{[18]} Bowick M. and Br\'ezin E., Phys. Lett. B, {\bf 268} (1991) 21.
\item{[19]} Gross D. J. and Klebanov I. R., Nucl. Phys., {\bf B352} (1991) 671.
\item{[20]} Dyson F. J., J. Math. Phys., {\bf 13} (1972) 90.
\item{[21]} Beenakker C. W. J., Phys. Rev. Lett., {\bf 70} (1993) 4126;
Beenakker C. W. J. and Rejaei B. (unpublished).
\item{[22]} Sutherland B., J. Math. Phys., {\bf 12} (1971) 246; {\bf 12}
(1971) 251; Phys. Rev. A, {\bf 4} (1971) 2019; {\bf 5} (1972) 1372.
\item{[23]} Bronk B. V., J. Math. Phys., {\bf 6} (1965) 228.
\item{[24]} Forrester P. J., Nucl. Phys., {\bf B402} (1993) 709; Tracy C. A.
and  Widom H., Commun. Math. Phys. (to appear).
\item{[25]} Szeg\"o G., {\it Orthogonal polynomials}, 3rd edition (American
Mathematical Society, Providence, RI) 1967.
\item{[26]} Hastings S. P. and McLeod J. B., Arch. Rat. Mech. Anal., {\bf 4}
(1980) 31.
\item{[27]} Kravtsov V. E. and Zirnbauer M. R., Phys. Rev. B, {\bf 46} (1992)
4332.

\bye